\documentstyle[aps,preprint]{revtex}
\draft
\begin{document}
\title{
    Quantum searching's underlying $SU(2)$ structure
    and its quantum decoherence effects
    }
\author{
    Sixia Yu and Chang-Pu Sun\cite{email,www}
    }
\address{
    The Institute of Theoretical Physics,
    Academia Sinica, Beijing 100080, P.R. China
    }
\date{November 17, 1998}
\maketitle
\begin{abstract}
The search operation for a marked state by means of Grover's
quantum searching algorithm is shown to be an element of group
$SU(2)$ which acts on a 2-dimensional space spanned by the marked
state and the unmarked collective state. Based on this underlying
structure, those exact bounds of the steps in various quantum
search algorithms are obtained in a quite concise way. This
reformulation of the quantum searching algorithm also enables a
detailed analysis of the decoherence effects caused by its
coupling with an environment. It turns out that the environment
will not only make the quantum search invalid in case of complete
decoherence, where the probability of finding the marked state is
unchanged, but also it may make the quantum search algorithm worse
than expected: It will decrease this probability when the
environment shows its quantum feature.
\end{abstract}

Since Shor \cite{shor} convincingly demonstrated in 1994 that
quantum mechanics can help in factoring a large number
exponentially faster than any known classical algorithms, there
were a few algorithms presented to overcome the classical limits
on the usual computation process \cite{ekert,steane}. Among them
the Grover's quantum searching algorithm \cite{gr1,gr2} has been
paid much attention \cite{lanl1,lanl2,lanl3} to at present as it
has been implemented by using nuclear magnetic resonance
techniques \cite{chung1}.

As well as in other quantum algorithms in quantum computation, the
non-classical feature such as quantum coherence plays a dominant
role in Grover's searching algorithm. But the environment
surrounding the qubits may force the quantum computer to become
classical by decohering the coherent superposition of quantum
states. This decoherence effect certainly makes the quantum algorithms
invalid \cite{chung2,palma} and thus the problem of decoherence
must be overcome or avoided before the quantum computation can be
implemented in practice. Recently, with Shor's factoring algorithm
as an explicit example, we have analyzed the decoherence problem
of quantum computation in detail \cite{sun1} based on the
generalization of the quantum dynamic theory of quantum
measurement [13-20]. The purpose of this note is to investigate
the decoherence influences on Grover's quantum algorithm. To this
end, a concise formulation of the quantum search algorithm is
presented using its $SU(2)$ structure at first. And then based on
this underlying structure we analyze the decoherence effects on
this algorithm in some details.

Classically, when there are $N$ objects among which there is an
unknown marked one, the best way to find this marked object is to
search for it one by one among those $N$ objects. For large $N$
the steps of classical search are therefore of order $O(N)$ in
order to find the marked object. By using the Grover's searching
algorithm \cite{gr1} however, the quantum computer needs only
$O(\sqrt N)$ steps of searching to find a marked quantum object
(e.g. state) with a probability near one.

Suppose that our system has $N$ levels and its Hilbert space is
spanned by mutual orthogonal bases $|k\rangle$ with
$k=1,2,\ldots,N$ and $N$ very large. Initially, the system is
prepared in a homogeneous coherent superposition
\begin{equation}\label{is1}
|\psi\rangle=\frac 1{\sqrt N} \sum_{k=1}^N |k\rangle
\end{equation}
of those $N$ bases. The problem is to find a certain marked basis
state, say $|m\rangle$ among those $N$ basis states in this
homogeneously superposed state.

According to Grover's searching algorithm, one step of search for
the marked state consists of the following two operations. The
first one is a selective phase rotation that is represented by
operator $I_m=1-2|m\rangle\langle m|$. It reverses the sign of the
coefficient before the marked state and keeps those phase factors
before other basis states unchanged. This calls for a quantum non
demolish measurement because one has to look for the marked state,
though with a very small probability, in order to change its
phase. The second operation is a diffusion transformation that is
represented by operator $-I_\psi=2|\psi\rangle\langle\psi|-1$. In
fact it is another selective phase rotation that keeps the phase
of the initial state unchanged and reverses the phase factors
before all those orthogonal states.

Before proceeding any further, we notice that together with the
marked state $|m\rangle$, a collective state defined by
\begin{equation}
|c\rangle=\frac1{\sqrt{N-1}} \sum_{k\ne m}^N|k\rangle
\end{equation}
spans a two-dimensional subspace ${\mathcal H}_S$ of the system.
The search operation is {\em in de facto} performed in this
subspace. This is because firstly, with notation
\begin{equation}
\sin\theta=\frac 1{\sqrt N},
\end{equation}
the initial state of the system
$|\psi\rangle$, rewritten as
$|\psi\rangle=\sin\theta|m\rangle+\cos\theta |c\rangle$
belongs to this subspace. And secondly those two operations $I_m$
and $I_\psi$ in one step of searching, since they are
determined only by states $|m\rangle$ and $|\psi\rangle$, belong
also to this subspace ${\mathcal H}_S$ and leave the subspace {\em
orthogonal} to ${\mathcal H}_S$ invariant.

On this observation and by denoting $\langle m|=(1,0)$ and
$\langle c|=(0,1)$ in ${\mathcal H}_S$,  one step of search
operation defined above can be expressed simply as
\begin{equation}
S(\theta)=-I_\psi I_m=\left(\matrix{\cos2\theta&\sin2\theta\cr
-\sin 2\theta&\cos2\theta}\right),
\end{equation}
which is exactly a rotation of $2\theta$ in the searching space
${\mathcal H}_S$, i.e., an element of group $SU(2)$ (not only
$SO(2)$ as will see later). When it acts on the subspace
orthogonal to ${\mathcal H}_S$, it changes only the signs of all
the bases in this subspace. This operation is in fact equivalent
to a time evolution $S(\theta)=e^{-it_0H_0}$ under Hamiltonian
\begin{equation}
H_0=i2\theta/t_0(|m\rangle\langle c|-|m\rangle\langle c|),
\end{equation}
where $t_0$ is the time interval required by a single search
operation. After $l$ searching steps, the resulting unitary
transformation $S^l(\theta)=S(l\theta)$ is a rotation of
$2l\theta$ and the probability of finding the system in the marked
state is therefore
\begin{equation}
P_G(m,l)=\left|\langle m|S(l\theta)|\psi\rangle\right|^2
      =\sin^2(2l+1)\theta.
\end{equation}
Since $\theta\approx 1/\sqrt N$ for large $N$, the optimum number
$l_G$ of searching steps is therefore the minimum maximizing
probability $P_G(m,l_G)=1$, which is given asymptotically by
\begin{equation}
l_G\approx \frac\pi4\sqrt N-\frac 12.
\end{equation}
This is exactly the tight bound for Grover's quantum search
algorithm \cite{lanl1} and shows that the searching steps are of
order $O(\sqrt N)$ in this quantum search algorithm.

Now let the system be initially in a general normalized state
$|\gamma\rangle$ and one tries to search for the marked state
$|m\rangle$, whose probability amplitude in the initial state
$\langle m|\gamma\rangle=\sin\theta_\gamma e^{i\phi}$ is small. As
will see immediately, this case is equivalent to Grover's search
algorithm using almost any unitary transformation \cite{gr2}.
Similarly, one step of search consists of two operations $I_m$ and
$I_\gamma=1-2|\gamma\rangle\langle\gamma|$. The quantum search
operation is then represented by $S_\gamma=-I_\gamma I_m$. By
introducing a normalized collective state
\begin{equation}
|c_\gamma\rangle=\frac{1+I_m}{2\cos\theta_\gamma}|\gamma\rangle,
\end{equation}
again we obtain a 2-dimensional searching space spanned by this
collective state and the marked state, where the search operation
is performed. The initial state of the system can be rewritten as
$|\gamma\rangle=\sin\theta_\gamma
e^{i\phi}|m\rangle+\cos\theta_\gamma|c_\gamma\rangle$ and the
searching operator as
\begin{equation}
S_\gamma=\left(\matrix{\cos2\theta_\gamma &\sin2\theta_\gamma
e^{i\phi}\cr -\sin2\theta_\gamma
e^{-i\phi}&\cos2\theta_\gamma}\right),
\end{equation}
which is truly an element in a group $SU(2)$ acts on the subspace
spanned by the collective state and the marked state. After
$l_\gamma$ searching steps and noticing
that $\theta_\gamma\approx|\langle m|\gamma\rangle|$ for small
amplitude $\langle m|\gamma\rangle$, the probability of finding the
marked state is maximized asymptotically by
\begin{equation}\label{lgma}
l_\gamma\approx \frac\pi{4|\langle m|\gamma\rangle|}-\frac 12.
\end{equation}
By identifying the marked state $|m\rangle$ with $|\tau\rangle$
in Ref.\cite{gr2}
and the initial state with $U|\gamma\rangle$ for an arbitrary
unitary transformation $U$, this gives the tight bound for
the so-called search algorithm using arbitrary unitary
transformation.

As another special case, it gives also the tight bound for the
search algorithm of finding the multiple quantum state, or to find
$s$ \lq\lq needles in the same haystack\rq\rq \cite{lanl1}.  In
this case the initial state is the same as $|\psi\rangle$ in
Eq.(\ref{is1}) while the marked state is itself a superposition
state $|m_s\rangle=\frac1{\sqrt s}\sum_{k=1}^s|k\rangle$. We have
therefore $\langle m_s|\psi\rangle={\sqrt{s/N}}$. From
Eq.(\ref{lgma}) we see clearly that the number of searching steps
in this case is approximately $l_s\approx \frac\pi4\sqrt{N/s}$.

The $SU(2)$ structure of the quantum search algorithm revealed
above provides us a convenient starting point to investigate the
influence of the quantum decoherence, which is caused by the
environment surrounding the qubits system that carries out the
quantum computation, on the quantum search algorithm. In what
follows we will make Grover's original searching algorithm as
an example to discuss decoherence effects of environment.

According to  the dynamical decoherence theory for quantum
computation developed recently in Ref.\cite{sun1}, the effects of
the environment on the computing process of Grover's quantum
search algorithm can be understood in the viewpoint  of state
entanglement or state correlation. Let's start from the initial
state $|\psi_T(0)\rangle=|\psi\rangle\otimes|e\rangle$ of the
total system formed by the qubits plus the environment with
$|e\rangle$ being the initial state of the environment. Suppose
that the time evolution of the system has a factorized structure
as $U_T(t)=S(l\theta)U_N$, where
$U_N|k\rangle\otimes|e\rangle=|k\rangle\otimes|e_k\rangle$ with
$|e_k\rangle=U_k|e\rangle$. After $l$ searching steps, the initial
state $|\psi_T(0)\rangle$ evolves into the following correlated
state
\begin{equation}
|\psi_T(t)\rangle=U_T(t)|\psi_T(0)\rangle
=\frac 1{\sqrt N}|u_m\rangle\otimes|e_m\rangle
 +\frac 1{\sqrt N}\sum_{k\ne m}^N|u_k\rangle\otimes|e_k\rangle
\end{equation}
at time $t=lt_0$, where $|u_j\rangle=S(l\theta)|j\rangle$
$(j=1,2,\ldots, N)$ reflects the free evolution without
environment. We see that each environment pointer state
$|e_j\rangle$ entangles with a qubit state $|u_j\rangle$. This
entanglement or correlation is essentially the physical origin of
the quantum decoherence.

In fact, at the presence of the environment, the \lq\lq
effective\rq\rq quantum state of the qubit system is no longer
pure due to the random disturbances of the environment. The
reduced density matrix is
\begin{eqnarray}
\rho={\rm Tr}_e(|\psi_T(t)\rangle\langle\psi_T(t)|)
    =\frac 1N\left(1+{\sum_{k\ne k^\prime}^N}^\prime|u_k\rangle\langle u_{k^\prime}|F_{k,k^\prime}+
      \sum^N_{k\ne m}(|u_m\rangle\langle
         u_k|F_{k,m}+{\rm H.c.})\right)
\end{eqnarray}
where ${\rm Tr}_e$ denotes the trace operation on the environment
variables and $\sum^\prime$ is a summation without index $m$ and
the so-called decoherence factor $F_{k,j}=\langle e_k|e_j\rangle$
characterizes the influence of the environment on the system.

By using the $SU(2)$ representation of the search operation
$S(l\theta)$, one have easily $\langle m|u_k\rangle=\sin\theta\sin
2l\theta$ for $k\ne m$ and $\langle m|u_m\rangle=\cos 2l\theta$.
After searching $l$ steps, the probability of finding the system
in the marked state $P(m,l)\equiv \langle m| \rho|m\rangle$ is
then given by
\begin{equation}
P(m,l)={\mathcal F}_1\sin^2(2l+1)\theta+(1-{\mathcal F}_1)\frac 1N
+\frac 12 ({\mathcal F}_2-{\mathcal F}_1)\sin 2\theta\sin 4l\theta
\end{equation}
where
\begin{equation}
{\mathcal F}_1
=\frac {\langle d|d\rangle-1}{N-2},\quad
{\mathcal F}_2=\frac {\langle d|e_m\rangle+\langle e_m|d\rangle}{2\sqrt{N-1}}
\end{equation}
with the collective environment state $|d\rangle$ defined by
\begin{equation}
|d\rangle=\frac 1{\sqrt{N-1}}\sum_{k\ne m}^N|e_k\rangle.
\end{equation}
We note that the norm of this collective state satisfies
$0\leq\langle d|d\rangle\leq N-1$. For example, when there is no
coupling between the environment and the system, i.e. all the
pointer states undergo the same time evolution, we have
$\langle d|d\rangle=N-1$; when all $|e_k\rangle$ are orthogonal we have
$\langle d|d\rangle=1$. Especially, when those pointer states
differ only in their phases, we can have $|d\rangle=0$. As a
result
\begin{equation}
\frac {-1}{N-2}\leq{\mathcal F}_1\leq 1, \quad |{\mathcal F}_2|\leq
1.
\end{equation}

If our system is decoupled from the environment, both two
decohering factors are one, i.e. ${\mathcal F}_1={\mathcal
F}_2=1$, and the coherence of the system is completely preserved in the quantum
computing process. In this case, the  optimum steps of searching
$l$ is determined by $l_G$ so that the probability of finding the
marked state is one.

Compared to the ideal case, the environment will cause a
decoherence when all the evolved states of the environment are
mutual orthogonal, i.e., ${\mathcal F}_1={\mathcal F}_2=0$. In
this case our system becomes essentially classical and the
quantum search operation tells us nothing about the marked state
because all the
off-diagonal elements of the reduced density matrix $\rho_c=1/N$
vanish completely and therefore the probability of finding the
system in the marked state
\begin{equation}
P_c(m,l)=\frac 1N
\end{equation}
remains to be the same as before the search operation.  This
result manifests that Grover's quantum searching algorithm will be
invalid when the quantum coherence is completely lost as expected.
The decoherence certainly destroys the effective quantum
computing. From another point of view, the quantum property of the
qubit system makes Grover's searching algorithm valid.

We can see more clearly the quantum feature of the Grover's search
algorithm by considering the interesting case when all the pointer
states are matched in their phases so that $|d\rangle=0$. In this
case ${\mathcal F}_1=-1/(N-2)$ reaches its minimum and ${\mathcal
F}_2=0$. The probability of finding the marked state is given by
\begin{equation}
P_n(m,l)=\frac{\cos^22l\theta}N.
\end{equation}
This probability is essentially zero (of order $o(1/N^2)$) when
$l=l_G$ as required by Grover's searching algorithm without
environment. This may be called a negative coherence, or {\it
necoherence}, effect of the environment. The quantum search
procedure in the presence of the quantum features of the
environment will be not only unable to find the marked state, it
may also make things worse than the classical search algorithm.
In this sense the fast search algorithm is at risk of being
ineffective than the classical case. The isolation of the qubit
system from the environment is more than necessary to make a
quantum search of the database.

For a more concrete example, since the quantum search operation
takes place only in a two-dimensional Hilbert space ${\mathcal H}_S$
spanned by states $|m\rangle$ and $|c\rangle$ or states
$|\pm\rangle=\frac 1{\sqrt 2}(|m\rangle \pm i|c\rangle)$, we
consider a minimum coupling between the system with the
environment given by Hamiltonian
\begin{equation}
H_T=H_0+H_+|+\rangle\langle+|+H_-|-\rangle\langle-|
\end{equation}
where $H_\pm$ depends only on observables of the environment.
At time $t=lt_0$, the system will be in the following state
\begin{equation}
|\psi_T(t)\rangle=e^{-itH_T}|\psi_T(0)\rangle=
e^{-i(2l+1)\theta}|+\rangle\otimes|e_+\rangle+
e^{i(2l+1)\theta}|-\rangle\otimes|e_-\rangle
\end{equation}
with $|e_\pm\rangle=e^{-itH_\pm}|e\rangle$. Denoting $\langle
e_+|e_-\rangle=re^{i2\delta}$, the probability of finding the
marked state is
\begin{equation}
P_r(m,l)=\frac 12(1-r)+r\sin^2\bigl((2l+1)\theta+\delta\bigr).
\end{equation}
When $r=1$ and
$\delta=\pi/2$, that is, the phase difference between two evolved
states are $\pi$, this probability becomes essentially zero
for the scheduled $l_G$ steps of searching. The necoherence effect
occurs again. This shows that Grover's search algorithm is
extremely sensitive to the phase change of the states of the
environment.

The quantitative value of $F_{k,k^\prime}$ depends on the details
of interaction between the qubit system and the environment, but
there still exists certain universality that enables us to discuss
the effect of decoherence further in some aspects. In the weak
coupling limit, any environment may be approximated by a bath of
harmonic oscillators \cite{x1,x2} and the interaction by a linear
coupling
\begin{equation}
H_I=\sum_{jk}(g_{jk}a_j|k\rangle\langle k|+{\rm H.c.)}
\end{equation}
where $g_{jk}$ is the coupling coefficients and $a_j$ are the
ladder operators for harmonic oscillators obeying
$[a_i,a_j^\dagger]=\delta_{ij}$. Assuming that there are
$M$ oscillators, we can explicitly obtain the
norm of the decohering factors at zero temperature \cite{palma,sun2}
\begin{equation}
|F_{k,k^\prime}|=\exp\left\{-\sum_{j=1}^M\frac{2(g_{jk}-g_{jk^\prime})^2}{\omega_J}\sin^2(\omega_jlt_0/2)\right\},
\end{equation}
where $\omega_j$  is the frequency of an oscillator in the bath.
For certain given spectral distribution, we can obtain the
decaying factor as
$|F_{k,k^\prime}|\approx e^{-\gamma_{k,k^\prime}lt_0}$
where $\gamma_{k,k^\prime}$ is the decoherence time
\cite{palma,sun2}. The effective steps $l$ in a fast quantum
searching are bounded therefore by the inverse of this decohering
time in the unit of $t_0$. Otherwise, this search can not be
finished before a needle in haystack is found.

To summarize, we recognized a subtle $SU(2)$ structure in Grover's
quantum search algorithm of fast quantum database searching and
its alternatives. With its help, we  discussed the influence of
the environment on these quantum computing processes and gave some
quantitative analysis for the probabilities of finding a marked
state of qubit surrounded by a random environment. To conclude
this note, we remark that firstly, the quantum coherence of qubits
characterized by the off-diagonal elements of its reduced density
matrix plays an important role in the computing process of
Grover's algorithm. Secondly, the negative coherence, {\em
necoherence}, effect caused by the phase matching of the pointer
states of the environment, though the off-diagonal elements do not
vanish, will make a greater hindrance to the implementation of the
quantum search algorithm than the quantum decoherence.

This work is partly supported by NFS of China.

\end{document}